\documentclass[authoryear,12pt]{elsarticle}

\usepackage{tikz,graphicx,lscape,subfig,amsmath,amssymb,natbib,lineno}

\numberwithin{equation}{section}

\begin{document}
\linenumbers
\title{A model of the number of antibiotic resistant bacteria in rivers}
\author[mdmu1]{Bonita Lawrence}
\author[mdmu2]{Anna Mummert\corref{cor1}}
\cortext[cor1]{Corresponding author. Email: mummerta@marshall.edu; Phone: (304) 696 3041; Fax: (304) 696 4646 }
\author[bdmu]{Charles Somerville}
\address[mdmu1]{Mathematics Department, Marshall University, Huntington, WV 25755, Email: lawrence@marshall.edu}
\address[mdmu2]{Mathematics Department, Marshall University, Huntington, WV 25755, Email: mummerta@marshall.edu}
\address[bdmu]{Biology Department, Marshall University, Huntington, WV 25755, Email: somervil@marshall.edu}

\begin{abstract}
The large reservoir of antibiotic resistant bacteria in raw and treated water supplies is a matter of public health concern.
Currently, the National Antimicrobial Resistance Monitoring Systems, a collaborative effort of the Centers for Disease Control, the US Department of Agriculture, and the US Food and Drug Administration, does not monitor antimicrobial resistance in surface waters.
Given the serious nature of antibiotic resistance in clinical settings, and the likelihood that antibiotic resistant bacteria can be transmitted to humans from large environmental reservoirs via drinking water, explanations for the distribution of antibiotic resistant bacteria and tools for studying this distribution must be found.
Here we focus on mathematical modeling of cultivable bacteria in a river, which will be used to study the distribution of antibiotic resistant bacteria in the environment.  We consider both antibiotic resistant and non-antibiotic resistant bacteria in the model, and, taking into account the strong correlation between land use and antibiotic resistant bacteria in rivers, we include a function for the influx of bacteria into the river from the shore.
We simulate the model for two different time scales and show that if there is too many bacteria from the land entering the river, the river entirely fills with antibiotic resistant bacteria, while less frequent influxes allows time for the bacteria to lose the antibiotic resistant gene.  This mathematically verifies that reduction in antibiotic use near the banks of rivers, will reduce the counts of antibiotic resistant bacteria in rivers.
\end{abstract}

\begin{keyword}
mathematical model \sep non-linear population dynamics \sep antibiotic resistance \sep bacteria \sep riverine system
\end{keyword}

\maketitle

\section{Introduction}

\label{introduction}

The age of antibiotics is usually traced back to 1928 - the year that Alexander Fleming discovered penicillin.  The first clinical uses of penicillin occurred in the early 1940's \citep{Dawson1941, Parascandola1997,Grossman2008}, and penicillin resistance was first reported in hospital isolates in 1947, only a few years after its introduction \citep{Lewis1995}.  Since that time, antibiotic resistance in clinically relevant bacteria has become a major health concern.  The US Food and Drug Administration estimates that approximately 70 percent of pathogenic bacteria encountered in hospital settings are resistant to one or more of the drugs commonly used in their treatment \citep{Bren2002}.  Concern over the spread of antibiotic resistant bacteria led to the formation of the National Antimicrobial Resistance Monitoring System (NARMS) in 1997 \citep{NARMS2006}.  NARMS is a collaborative effort of the Centers for Disease Control, the US Department of Agriculture, and the US Food and Drug Administration.  Although NARMS monitors antibiotic resistant bacteria isolated from humans, food animals, and raw food products, it does not monitor antimicrobial resistance in surface waters.

The nation's surface waters are vital, multiple-use resources.  Major rivers, in particular, are used as transportation routes, recreational venues, and industrial water sources, as well as, receptacles of stormwater runoff, sanitary sewage, and industrial wastes.  In addition, surface waters are the sources of drinking water for millions of people worldwide.  For example, the Ohio River Valley Water Sanitation Commission  estimates that the Ohio River provides drinking water for more than five million people \citep{ORSANCO}.  Several studies have shown that large numbers of freshwater bacteria found in these life sustaining waterways are resistant to one or more antibiotics \citep{Cooke1976,Ash2002,Niemi1983,Kelch1978,Bennett1999,Smith2006,Dotson2008}.  The resistance found in freshwater bacteria is problematic for humans since, it has been shown that antibiotic resistant bacteria can be isolated from tap water \citep{Armstrong1981,Nagy2009}.  Therefore, the large reservoir of antibiotic resistant bacteria in raw and treated water supplies is a matter of public health concern.

Many studies have correlated the presence of antibiotic resistant bacteria in surface waters with the human activities occurring along the shore.  Land uses that have been correlated with antibiotic resistance include wastewater treatment plants \citep{Frank2005}, urban areas and other impervious surfaces \citep{Boon1999,Somerville2004}, industrial and heavy metal pollution \citep{McArthur2000}, and agricultural pastures and row crops \citep{Kuhn2005,Dotson2008}.  As an example, regarding the Mud River, WV, Dotson specifically studied the watershed land use in relation to the count and distribution of antibiotic resistant bacteria, and concludes ``the evidence shows a correlation of antibiotic resistant bacteria to areas of livestock in the watershed~\citep{Dotson2008}."

The source of environmental antibiotic resistant strains has been attributed to the use of antibiotics both in medical and agricultural settings.  Therefore, previous studies have assumed that aquatic antibiotic resistance derives directly from human and animal fecal contamination entering surface waters via inadequate sewage treatment or runoff \citep{Cooke1976,Ash2002,Raloff1999}. If this assumption holds true, then antibiotic resistant bacteria in an environmental sample should be a subset of fecal bacteria, and their distribution should be predicted using standard methods for the microbiological assessment of water and wastewater.  However, recent studies have shown that cultivable antibiotic resistant bacteria are not a subset of fecal indicator bacteria, which implies that their distribution in freshwater environments cannot be predicted by standard techniques \citep{Smith2006,Somerville2007,Dotson2008}.

We conclude, there is a large population of antibiotic resistant bacteria in surface waters and drinking water that are not monitored under NARMS surveillance, and whose distribution cannot be predicted by standard water quality analyses.  Given the serious nature of antibiotic resistance in clinical settings, and the likelihood that antibiotic resistant bacteria can be transmitted to humans from large environmental reservoirs via drinking water, tools for modeling the distribution of antibiotic resistant bacteria in the environment are needed.

\cite{Mahloch1974} offers one of the first comparisons of mathematical models of bacteria in rivers.  Mahloch compares six models for the count and distribution of coliform bacteria in rivers, three deterministic and three stochastic, with data collected from the Leaf River, MS.  The six models do consider influx of bacteria from tributaries, though not from other sources, such as runoff from agricultural pastures and row crops.  Also, the models are for counts of bacteria only, not for bacteria with or without an antibiotic resistant gene.

Since Mahloch, models have been introduced to study the transfer of the antibiotic resistant gene, both at the bacteria level and at the human level.  The mechanisms of gene transfer have been modeled with a mass action term ~\citep{Levin1979} and later with a Michaelis-Menten kinetic \citep{Andrup1998}.  Transfer of antibiotic resistance among humans is most notable in hospitals.  In the hospital setting, \cite{Webb2005} present a model which captures both the patient level and the bacteria level, in the sense of the bacterial load contributed by each patient.  We know of no models, of the sort we present here, that focus on antibiotic resistant bacteria in a river with the possibility of bacteria entering the river due to human activity along the shore line.

In this paper we present a novel model for cultivable bacteria in a river focusing on the influx of bacteria due to human activities along the shore.  We consider both antibiotic resistant and non-antibiotic resistant bacteria in the model.  Taking into account the strong correlation between land use and antibiotic resistant bacteria in water, we include a function for the influx of bacteria into the river from the shore.  Our model includes the possibility of both vertical and horizontal transfer of the antibiotic resistant gene, and the fact that the land bacteria are not adapted to the river ecosystem and will not survive in the river.
In Section~\ref{themodel} we describe the model and the model implementation.  Realistic parameters are discussed and the model is simulated in  Section~\ref{simulation}.  We conclude with our discussion in Section~\ref{discussion}.

\section{Model assumptions and implementation}
\label{themodel}

Throughout our model description we use the phrase ``antibiotic resistant'' to mean resistance in the clinical sense to a particular unspecified antibiotic, such as, tetracycline, and ``non-antibiotic resistant'' to mean not resistant to that particular antibiotic.  In general, any one bacterium could be non-resistant to all antibiotics, or resistant to any number of antibiotics at once.  Though we consider only clinical resistance to one particular antibiotic, this model could easily be extended to consider bacteria with multiple resistance or to compare clinical and weaker forms of resistance.

We use the independent variable time~$t$ in our model to represent time from the head of the river.  As time passes, we assume the bacteria are transported down the river in such a way that all bacteria in the same cohort at time~$t$ stay together as they travel down the river.  Typical data collected from rivers (e.g.~\cite{Dotson2008,Smith2006}) mark the water sample collection points using mile marks along the direction of the river, not time, which leads one to believe that distance should be the independent variable.  However, typically mathematical models use time as the independent variable, as we do here.  This should cause the reader no concern, since by using river flow rates, distance from the head of the river can be converted to time from the head of the river.

Only some bacteria are able to be cultivated and counted using current methods, and so, throughout, we consider only cultivable bacteria in the river.  In general, there are many more bacteria in a river that cannot be cultivated and counted.  The main model assumptions are schematically represented in Figure~\ref{ModelPicture}.  Details of the model assumptions and their mathematical implementation are presented below.

We consider two distinct classes of cultivable bacteria in the river: those that are ``river" bacteria,~$R$, and those that are ``land" bacteria,~$L$.
The river bacteria are naturally occurring bacteria which are adapted to survive in the river.  The land bacteria enter the river from the shore and we assume they are not adaptable to the river. Mathematically, the land bacteria have an additional death rate, which is large enough to offset any reproduction.  Therefore, the only increase in the land bacteria in the river comes in the form of immigration from the shore.
In the river, all bacteria are homogeneously mixed, so any bacterium can come into contact with any other bacterium.
We assume that neither type of bacteria can transform into the other.

Both the river and land bacteria are further subdivided into those that have the antibiotic resistant gene, called ``resistant'',~$R_{I}, L_{I}$, and those that do not, called ``susceptible'',~$R_S, L_S$.  We always have one particular gene in mine, for example, the gene to make a bacterium resistant to tetracycline.
In this model, we are most interested in assessing human activities along the shore in relation to the antibiotic resistance in rivers, so we assume the land bacteria which are antibiotic resistant gained the gene through human activity, though some may naturally be resistant (see for example~\cite{DCosta2006}).

The bacteria with the antibiotic resistant gene can transfer the gene to the susceptible bacteria and the bacteria with the antibiotic resistant gene can lose the gene.  A resistant bacterium of either type can transfer the gene to a susceptible bacterium of either type.  The transfer of the antibiotic resistant gene
is similar to the transfer of a disease: one `infected' (resistant) bacterium comes into contact with a `susceptible' bacterium and with some probability the result of the contact is two `infected' (resistant) bacteria.  In particular, a resistant bacterium does not lose the antibiotic resistant gene when the gene is transferred.
An susceptible-infected (SI) disease model is used to represent the transfer of the antibiotic resistant gene.
\cite{Gonzalo1989} shows that antibiotic resistant bacteria survive less in less polluted water.  Therefore, we assume that the loss of the antibiotic resistant gene depends on the pollution, $P$, in the river.  The loss rate is $B(P)$, where the function has the property that as $P$ increases $B$ decreases and as $P$ decreases $B$ increases.  Many such functions exists, and we use in this model $\displaystyle{B(P) = \frac{\beta}{L_S + L_I +1}}$, so that pollution is measured by the bacteria in the river from the shore.

Both the river bacteria and the land bacteria compete for space resources in the river leading to a river carrying capacity.
The carrying capacity for the river used in the model is the carrying capacity for only the cultivable bacteria; in general, a river will have a larger carrying capacity for all bacteria including those that are not cultivable.
We assume that the river has a constant carrying capacity for the length relevant to these interactions.
A logistic growth model is used to represent the populations' dependence on the river carrying capacity.  The river carrying capacity can allow an increase in the number of river bacteria, but not in the number of land bacteria.  We assume that the additional death rate of the land bacteria is large enough to offset any population growth that occurs due to the river carrying capacity.

Since we assume that neither river nor land bacteria can transform into the other, the influence of the land bacteria (river bacteria) on the river bacteria (land bacteria) occurs in two ways: through contact and transfer of the antibiotic resistant gene, and through total population size in accordance with the river carrying capacity.

Land bacteria can enter the river at time~$t$.  The increase in land bacteria in the river is split into those with the antibiotic resistant gene and those without,~$F_I(t)$ and~$F_S(t)$, respectively.  We assume that $F_I(t) \geq 0$ and $F_S(t) \geq 0$ for all $t$.  Generally, these rates are not constant and depend on time (location down river).  Mathematically, these terms provide an external forcing term for the model.

For simplicity, we assume that all rates which effect both river and land bacteria are constant and are the same for both types of bacteria. For example, the antibiotic resistance gene transfer rate is a constant~$\alpha$ for both types of bacteria.

Let the total cultivable bacteria in the river be denoted $TCB = R_S + R_I + L_S + L_I$.  Note that~$TCB$ is not a constant since bacteria are added to the river from the land.  Set the pollution measure to be $L = L_S + L_I$.  Our model can be described with the following system of differential equations.
\begin{eqnarray}
\label{themodelequations}
\frac{dR_S}{dt} & = & -\alpha R_S(R_I+L_I) +
\left(\frac{\beta}{L +1}\right) R_I +r\left( 1 - \frac{TCB}{K} \right)R_S \notag\\
\frac{dR_I}{dt} & = & \alpha R_S(R_I+L_I) - \left(\frac{\beta}{L +1}\right) R_I +r\left( 1 - \frac{TCB}{K} \right)R_I \\
\frac{dL_S}{dt} & = & F_S(t)-\gamma L_S-\alpha L_S(R_I+L_I) + \left(\frac{\beta}{L +1}\right) L_I +r\left( 1 - \frac{TCB}{K} \right)L_S \notag\\
\frac{dL_I}{dt} & = & F_I(t)-\gamma L_I +\alpha L_S(R_I+L_I) - \left(\frac{\beta}{L +1}\right) L_I +r\left( 1 - \frac{TCB}{K} \right)L_I \notag
\end{eqnarray}

The parameter interpretations are:~$\alpha$ is the transfer rate of the antibiotic resistant gene;~$\beta$ is the loss rate of the antibiotic resistant gene;~$r$ is the demographic rate due to the river carrying capacity;~$F_S(t)$ and~$F_I(t)$ are the rates of bacteria entering the river from the shore; and~$\gamma$ is the death rate due to the land bacteria not being adapted to survive in the river.  The parameter interpretations, along with their dimensions, are also presented in Table~\ref{parametertable}. (In Table~\ref{parametertable}, AR is used as an abbreviation for antibiotic resistant).  We assume that the land bacteria death rate~$\gamma$ is large enough to offset any population growth that occurs due to the river carrying capacity; specifically, when $\gamma > r$ any reproduction, mathematically due to the carrying capacity term, will be offset by the bacterial death due to not being adapted to survive in the river.

It is easy to verify that the non-negative state space is invariant, that is, any initial condition (a set of four population levels) that is non-negative results in population levels that are non-negative for all time.

\section{Model Simulation}
\label{simulation}
The mathematical model presented here can qualitatively and quantitatively describe the distribution of antibiotic resistant bacteria in a river effected by bacteria entering the river from the shore, and the subsequent transfer and loss of the antibiotic resistant gene.

We demonstrate the model with a simulation of antibiotic resistant and non-antibiotic resistant land bacteria entering a river which initially has no antibiotic resistant bacteria.  Most model parameters are determined from literature on survival and reproduction of bacteria in rivers, and on transfer and loss of antibiotic resistant genes in rivers.  The parameter values used in our simulations and their corresponding references are given Table~\ref{parametervalues}.  For parameters $r$ and $\gamma$ the values were determined using information on the half-life of bacteria in rivers found in  \cite{Hendricks1972} and the model developed for the Ohio River by \cite{LimnoTech}, respectively.  The transfer rate $\alpha$ applies in situations when the water (river specifically) is dilute; it was taken from \cite{Andrup1999} who assume that time is measured in hours.  The antibiotic resistance loss rate is known to be higher in less polluted water \citep{Gonzalo1989}.  Thus we set the loss rate to be a function of pollution, in the model pollution is represented by the land bacteria.  If the pollution reaches 0, then the loss rate is $\beta = 100$, while if the pollution is at the river carrying capacity, then the loss rate is very small, $0.0001$.  We assume a river carrying capacity of 1,000,000.  Before the influx of any land bacteria, the river is assume to contain only non-antibiotic resistant bacteria and these bacteria have reached the carrying capacity, that is, $R_S(0)=K$, $R_I(0)=0$, $L_S(0)=0$, and $L_I(0)=0$.

To simulate the influx of land bacteria into the river, we assume that the bacteria enter the river from the shore at four distinct times, evenly spaced over the length of the river.  At each of these influx times, the rate of influx of antibiotic resistant bacteria was chosen randomly to be between 0 and 25.  The rate of influx of non-antibiotic resistant bacteria was also chosen randomly to be between 0 and 50, with the condition that the rate of influx of non-antibiotic resistant bacteria is larger than the influx of antibiotic resistant bacteria.  The fact that $F_S > F_I$ corresponds with data collected from the Mud River, WV~\citep{Dotson2008}.  At all other integer points, the influx is set to zero. The functions $F_S(t)$ and $F_I(t)$ are linear interpolations of these point sources.

Two different simulations are presented in Figures~\ref{Figure100} and~\ref{Figure1000}.  In both of these simulations, the same bacteria influx functions are used, while two different time intervals are considered.  The first has total time 100 and the second has time 1,000. In the first case, there is too much pollution and, if the simulation were to continue, all of the bacteria in the river will eventually have the antibiotic resistance gene.  Most of the bacteria will be river bacteria, $R_I$, though some will be land bacteria, $L_I$.  In the second case, the river has enough time between the influx times for most antibiotic resistant bacteria to lose the antibiotic resistant gene.  Between each influx, the river recovers and most bacteria are non-antibiotic resistant river bacteria, $R_S$.

Using these parameters and influx functions, and both of the time scales, a local parameter sensitivity analysis was performed for the model.  Each parameter was adjusted by $\pm 20\%$ (the functions $F_S$ and $F_I$ were considered together as one parameter and were adjusted up or down together) and the effect of this change was studied graphically.  For all populations and both time scales, the river carrying capacity rate, $r$, is the least influential parameter of the model, having no effect on the population levels throughout the entire simulation(s).  Each of the other parameters, $\alpha$, $\beta$, $K$, $\gamma$, and the functions $F_S$ and $F_I$, was more influential on the longer time scale 1000 than on the shorter time scale 100.

For the shorter time scale, the number of non-antibiotic resistant river bacteria, $R_S$, was overall the least influenced population, the population tends to 0 in every case, while the number of non-antibiotic resistant land bacteria was the most influenced population.  Figure~\ref{SenGraph100} shows the effect of the parameter sensitivity analysis on the non-antibiotic resistant land bacteria, $L_S$, on the shorter time scale. For the longer time scale, $R_S$, $R_I$, and $L_S$ were very influenced by changes in all parameters, except $r$, while $L_I$ was only influenced by changes in $\gamma$ and the influx functions.  Figure~\ref{SenGraph1000} shows the effect of the parameter sensitivity analysis on the non-antibiotic resistant river bacteria, $R_S$, on the longer time scale.

\section{Discussion}
\label{discussion}

The model presented here is, by its very definition, a generalization of the actual processes that effect the amount and distribution of antibiotic resistant cultivable bacteria in rivers.  We focus here on three aspects of antibiotic resistant bacteria in rivers: the influx of bacteria into the river from the shore, the transfer and loss of the antibiotic resistant gene, and the river carrying capacity.  The model presented captures these aspects well and, by making some simplifying assumptions, the model uses only five parameters and the unknown functions $F_S$ and $F_I$.

The model simulations demonstrate that human involvement can have a significant influence on the level of antibiotic resistance of bacteria in rivers.  After each of the land bacteria influx points, the level of antibiotic resistant river bacteria increases, while the non-antibiotic resistant river bacteria decreases. And, only with sufficient time between influxes, can the bacteria lose the antibiotic resistant gene and the river recover.  Sources of antibiotic resistance entering rivers must be determined and the corresponding activity adjusted to prevent the resistance gene from becoming typical in a river.  Without such interventions, antibiotic resistance will spread and clinical uses of antibiotics may become ineffective, leading to serious public heath crises.

There are aspects of the transfer of the antibiotic resistant gene between bacteria that are not captured by this model.  For example, we assume (implicitly) that when a bacterium `dies' it can no longer transmit the antibiotic resistant gene to another bacterium.
In reality, some bacterial death can cause pieces of a bacterium's cell makeup to disperse into the river, where it is possible for some other organisms to collect these dispersed pieces, including the antibiotic resistant gene.   For another example, some researchers distinguish between antibiotic resistant (in the clinical sense) and weakly antibiotic resistant; this distinction does not appear in the model presented here, though extra population classes would be easy to add to the model for to separate clinical resistance from weakly resistant.
Despite its limitations, we believe that the model captures some crucial aspects regarding the dynamics of antibiotic resistant bacteria in rivers, specifically involving human use of antibiotics near rivers.

In this paper, we offer a model which is only a preliminary formalism of the river and antibiotic resistance situation, however, we believe that the main aspects of the situation have been accounted for in this model.  This novel model opens a wide range of possible future directions for ecological study.  The model and model simulations suggest some insights and refinements of the model, which will lead to better understanding of the count and distribution of antibiotic resistant bacteria in rivers.

\begin{enumerate}
 \item Due to the uncertainty of the influx of the land bacteria and the sensitivity of the model to the influx functions~$F_S(t)$ and~$F_I(t)$, care must be taken when these functions are determined.  Studies have shown that activity along the shore, such as farming row crops, livestock pastures, and the presence of urban areas, has a strong positive correlation with the amount of antibiotic resistant bacteria are in the river \citep{Dotson2008}.  Assessing this dependence and using the relationship to generate accurate influx functions needs to be addressed.

\item Most rivers changes size, both width across and depth, throughout their length, which leads to the conclusion that the carrying capacity should be a function of the length of the river, time t in this model.  For a given river, some measure of carrying capacity must be found and measured along the length of the river, and used to create a river carrying capacity function K(t) that depends on location along the river.

\item One obvious refinement would be to include spatial aspects in the model.  Here we assume, basically, that the river is one dimensional; it just has a length, which corresponds to time in the model.  In the river, bacteria homogeneously mix, and all bacteria in the cohort at time t travel down the river together.  From the shore, the land bacteria immediately enter the river and mix with the bacteria already present.  A more realistic model would include the spatial spread of the bacteria from the shore, and the possibility of spreading down the river at different rates, not traveling together as a cohort.

\end{enumerate}

We assert that this model could be generalized to any situation with non-native organisms entering an environment that they are not adapted to live in and their subsequent interactions with the native organisms.  The transfer and loss mechanism could apply to many genetic traits, including, possibly, those responsible for reduction and oxidation of metals.  The main environmental requirement that is assumed in this model for the interactions is homogeneous mixing of the organisms.  Any environment that ensured homogeneous mixing, would be suitable for this model.

\section{References}
\bibliographystyle{elsarticle-harv}
\bibliography{ARBModel}

\newpage
\begin{figure}
\centering
\usetikzlibrary{decorations.pathmorphing}
\usetikzlibrary{arrows,calc}
\usetikzlibrary{positioning}
\tikzset{>=stealth',
from shore/.style={dotted,->},
transfer/.style={<->,thick},
influence/.style={thick,dashed,->},
river edge/.style={decorate,decoration=snake},
influence2/.style={thick,dashed,bend right,->},}
\begin{tikzpicture}
\tikzstyle{shore}=[circle,thick,draw=black!75,fill=black!20]
\tikzstyle{river}=[circle,thick,draw=black!75]
\tikzstyle{location}=[rectangle,thick,draw=black!75]
\node[shore] (lishore) {$L_I$};
\node[shore] (lsshore) [above=2cm of lishore] {$L_S$};
\node[location] (shore) [above=of lsshore] {Shore};
\node (lowerleftriver) [below right=.9cm of lishore] {};
\node (placeholder1) [right= 4mm of lishore] {};
\node (placeholder2) [right=4mm of lsshore] {};
\node (placeholder3) [right=4mm of shore] {};
\node (upperleftriver) [above=of placeholder3] {};
\node[river] (liriver) [right=of placeholder1] {$L_I$};
\node[river] (lsriver) [right=of placeholder2] {$L_S$};
\node (placeholder4) [right=of placeholder3] {};
\node[location] (river) [right=of placeholder4] {River};
\node (placeholder5) [right=of lsriver] {};
\node[river] (ririver) [right=of placeholder5] {$R_I$};
\node[river] (rsriver) [below=2cm of ririver] {$R_S$};
\node (placeholder3) [right=of river] {};
\node (upperrightriver) [above right=of placeholder3] {};
\node (lowerrightriver) [below right=1.3cm of rsriver] {};
\node (lsdeath) [above left=6mm of lsriver] {};
\node (lideath) [below left=6mm of liriver] {};
\node (rsdeath) [below left=6mm of rsriver] {};
\node (rideath) [above left=6mm of ririver] {};
\draw[river edge] (upperleftriver) -- (lowerleftriver);
\draw[river edge] (upperrightriver) -- (lowerrightriver);
\draw[from shore] (lsshore) -- (lsriver);
\draw[from shore] (lishore) -- (liriver);
\draw[transfer] (lsriver) -- (liriver);
\draw[influence] (ririver) -- (lsriver);
\draw[transfer] (ririver) -- (rsriver);
\draw[influence] (liriver) -- (rsriver);
\path (liriver) edge [bend right, thick, dashed,->] (lsriver);
\path (ririver) edge [bend right, thick, dashed,->] (rsriver);
\path (ririver) edge [loop above,thick] (ririver);
\path (rsriver) edge [loop below,thick] (rsriver);
\path (lsriver) edge [thick, ->] (lsdeath);
\path (liriver) edge [thick, ->] (lideath);
\path (rsriver) edge [thick, ->] (rsdeath);
\path (ririver) edge [thick, ->] (rideath);
\end{tikzpicture}
\caption{Schematic representation of the model assumptions.  The dotted lines indicate bacteria entering the river from the shore.  The dashed lines indicate interactions between bacteria which cause a transfer of the antibiotic resistant gene.  The solid lines indicate the transfer or loss of the antibiotic resistant gene, reproduction, and death.  The land bacteria are not adapted to the river and we assume they never reproduce.  The river bacteria reproduce or die due to the influence of the river carrying capacity.  Details are provided throughout the main text.}
\label{ModelPicture}
\end{figure}
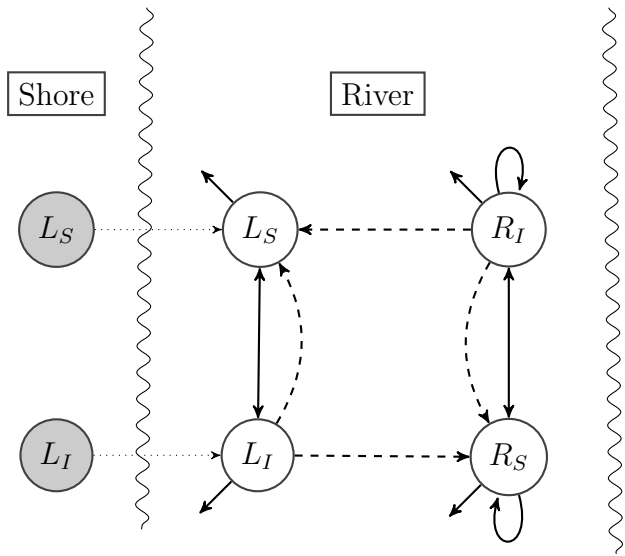

\begin{table}
\begin{center}
\begin{tabular}{lll}
\hline
Parameter & Description & Dimension \\ \hline
t & time from head of the river  & time \\[1mm]
$R_S$ & non-AR river bacteria  &
concentration\\[1mm]
$R_I$ & AR river bacteria  & concentration\\[1mm]
$L_S$ & non-AR land bacteria  &
concentration\\[1mm]
$L_I$ & AR land bacteria  & concentration\\[1mm]
$TCB$ & $R_S + R_I + L_S + L_I$ & concentration \\ [1mm]
$\alpha$ & transmission rate of AR gene & time$^{-1}$ $\times$ concentration$^{-1}$\\[1mm]
$\beta$ & loss rate of AR gene & concentration $\times$ time$^{-1}$\\[1mm]
$K$ & carrying capacity of river & concentration\\[1mm]
$r$ & birth-death rate due to $K$ & time$^{-1}$\\[1mm]
$F_S$ & entry of non-AR bacteria from land & concentration $\times$ time$^{-1}$\\[1mm]
$F_I$ & entry of AR bacteria from land & concentration $\times$ time$^{-1}$\\[1mm]
$\gamma$ & death rate of land bacteria & time$^{-1}$\\
\hline
\end{tabular}
\end{center}
\caption{Parameters, descriptions, and dimensions}
\label{parametertable}
\end{table}

\begin{table}
\begin{center}
\begin{tabular}{lrl} \hline
Parameter & Value & Reference \\\hline
$\alpha$ & 0.00006 & \cite{Andrup1999}\\
$\beta$ & 100 & \\
$K$ & 1,000,000 & \\
$r$ & 0.01& \cite{Hendricks1972}\\
$\gamma$ & 0.02 & \cite{LimnoTech}\\
 \hline
\end{tabular}
\end{center}
\caption{Parameter values and references.  Parameters without a reference were fixed from a reasonable range of possibilities for the simulations.}
\label{parametervalues}
\end{table}

\begin{figure}
\begin{center}
\includegraphics[scale=.65,clip, trim = .8in 1.85in .7in 1.7in]{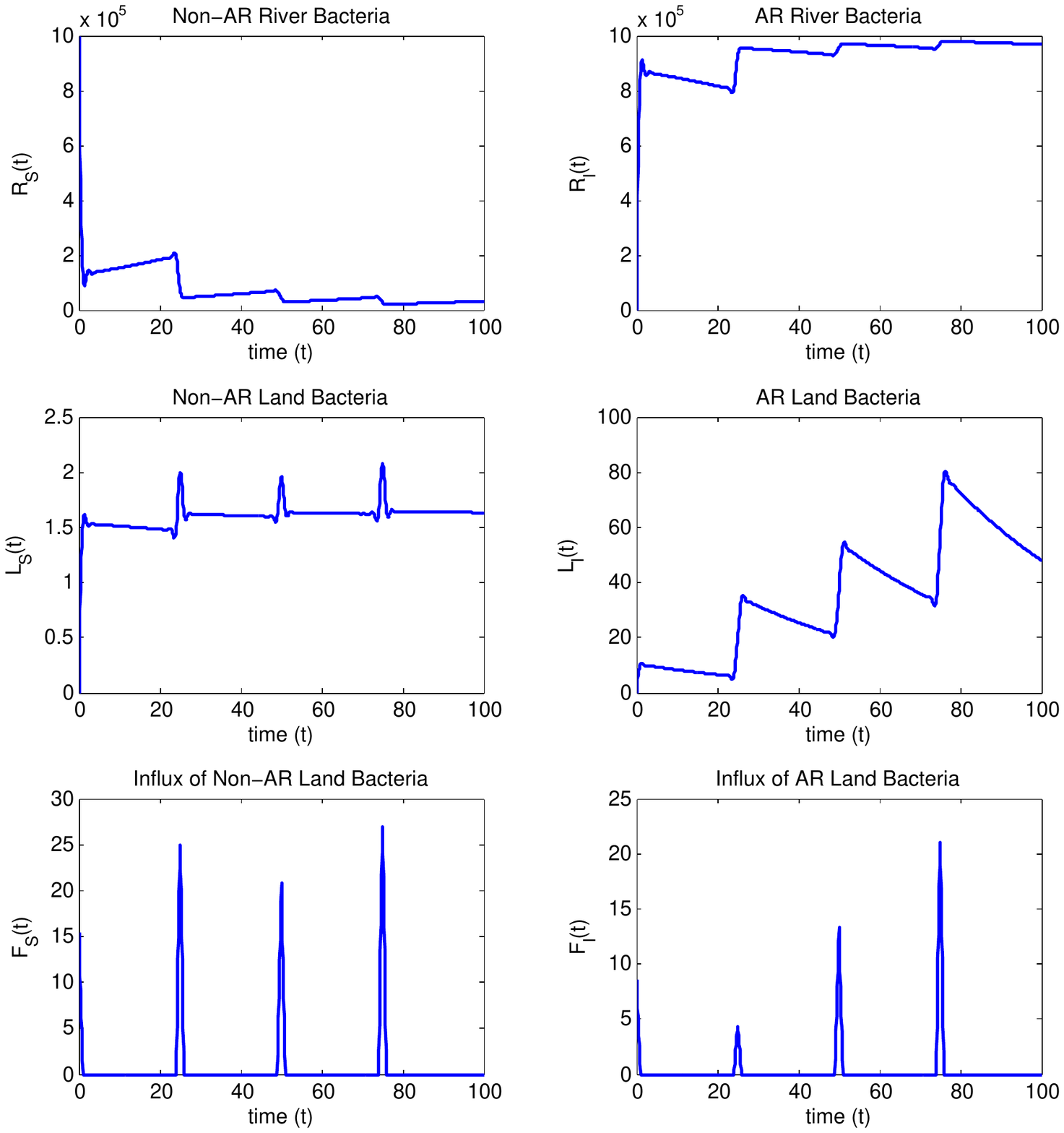}
\caption{Simulation of the populations of bacteria in a river with influx of bacteria from the land, over the time interval 0 to 100.}
\label{Figure100}
\end{center}
\end{figure}

\begin{figure}
\begin{center}
\includegraphics[scale=.65,clip, trim = .8in 1.85in .7in 1.7in]{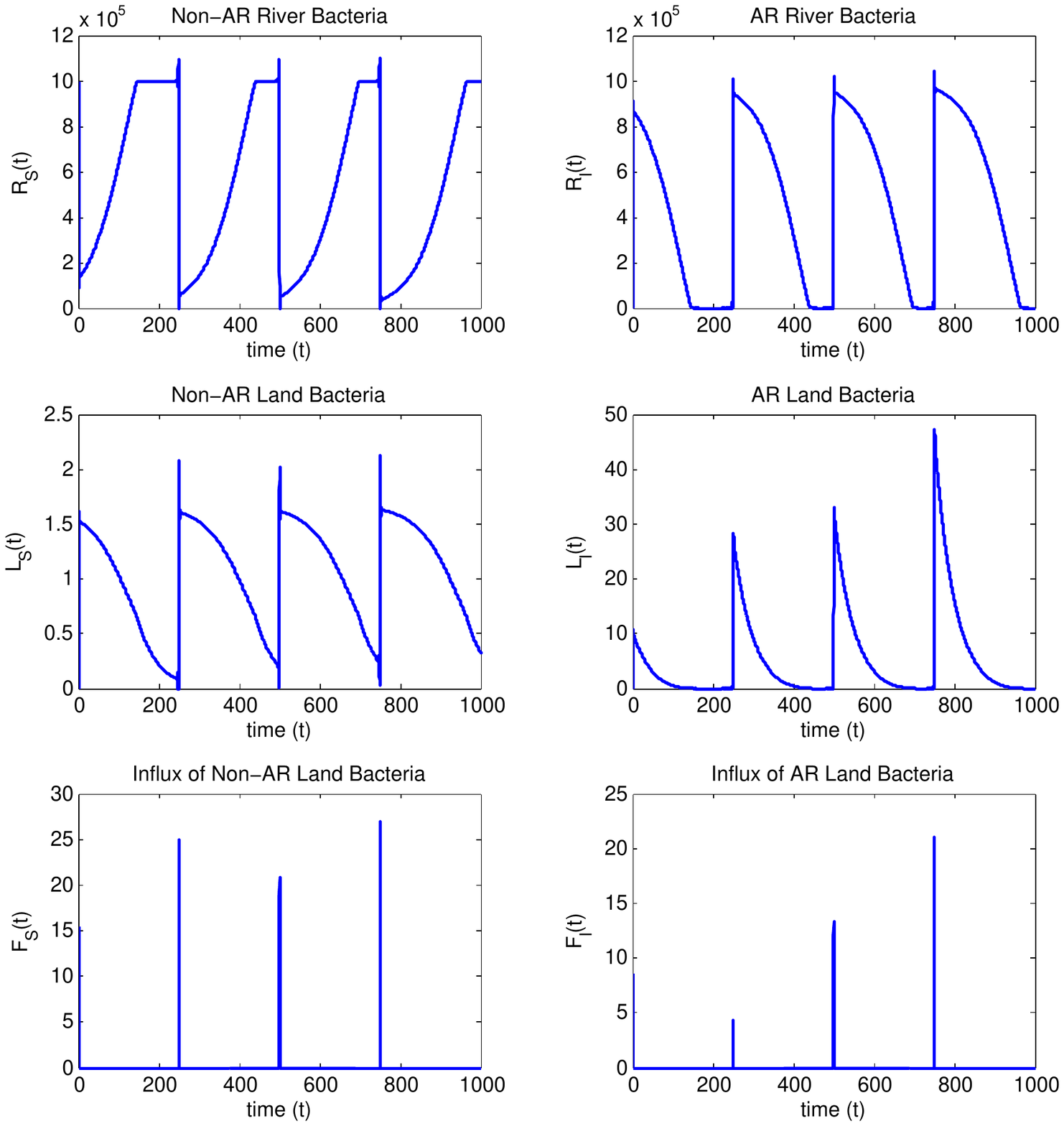}
\caption{Simulation of the populations of bacteria in a river with influx of bacteria from the land, over the time interval 0 to 1000.}
\label{Figure1000}
\end{center}
\end{figure}

\begin{figure}
\begin{center}
\includegraphics[scale=.65,clip, trim = .8in 1.85in .7in 1.75in]{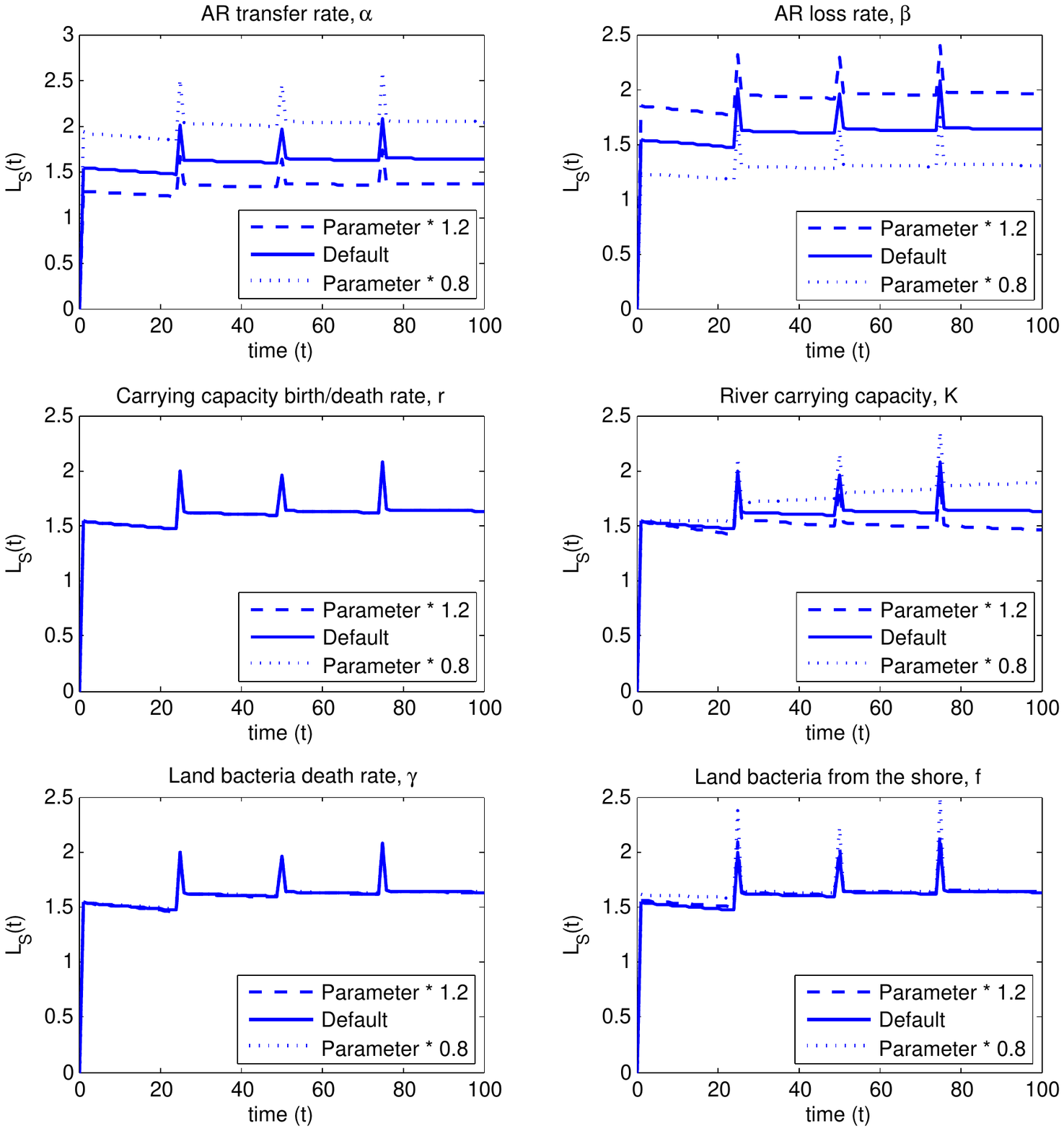}
\caption{Parameter sensitivities - effect on non-antibiotic land bacteria, $L_S$. Solid curve corresponds to the default parameter values; dashed curve corresponds to an increase in the parameter value by a factor of 20\%, while keeping all other parameters at their default value; dotted curve corresponds to a decrease in the parameter value by a factor of 20\%, while keeping all other parameters at their default value.}
\label{SenGraph100}
\end{center}
\end{figure}

\begin{figure}
\begin{center}
\includegraphics[scale=.65,clip, trim = .8in 1.85in .7in 1.7in]{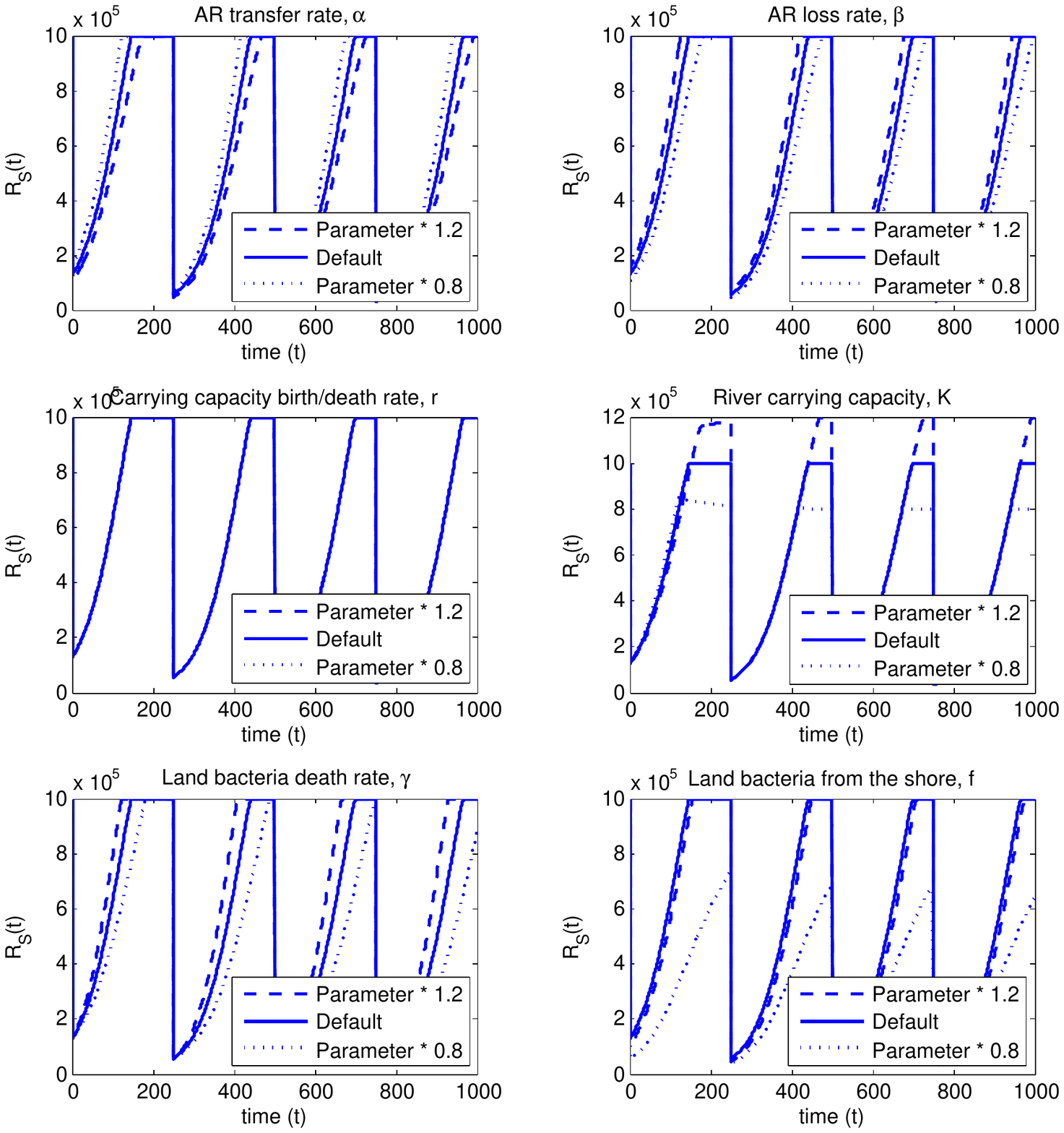}
\caption{Parameter sensitivities - effect on non-antibiotic river bacteria, $R_S$. Solid curve corresponds to the default parameter values; dashed curve corresponds to an increase in the parameter value by a factor of 20\%, while keeping all other parameters at their default value; dotted curve corresponds to a decrease in the parameter value by a factor of 20\%, while keeping all other parameters at their default value.}
\label{SenGraph1000}
\end{center}
\end{figure}

\end{document}